\documentclass[letterpaper,twocolumn,10pt]{article}
\usepackage{usenix,epsfig,bytefield}
\usepackage{ifthen}
\usepackage{xcolor}
\usepackage{booktabs}
\usepackage{color}
\usepackage{colortbl}
\usepackage{float}                           
\usepackage{bigdelim}

\usepackage[hyphens]{url}
\usepackage[colorlinks,citecolor=blue,breaklinks=true]{hyperref}%
\usepackage{svg}
\usepackage{lipsum}

\begin{document}

\newcommand\hmm[1]{\ifnum\ifhmode\spacefactor\else2000\fi>1000 \uppercase{#1}\else#1\fi}

\newcommand{\System}{LayerZero}
\newcommand{\MP}{OMP}
\newcommand{\Executor}{Executor}
\newcommand{\executor}{\hmm{e}xecutor}
\newcommand{\Verifier}{DVN}
\newcommand{\verifier}{DVN}

\newcommand{\showComments}{yes}
\newcommand{\note}[2]{
    \ifthenelse{\equal{\showComments}{yes}}{\textcolor{#1}{#2}}{}
}
\newcommand{\nt}[1]{\note{blue}{Note: #1}}
\newcommand{\step}[1]{\raisebox{.5pt}{\textcircled{\raisebox{-.9pt} {#1}}}}

\widowpenalty10000
\clubpenalty10000

\date{}

\title{\Large \bf \System{}}

{\author{\normalfont{Ryan Zarick \hspace{1em} Bryan Pellegrino \hspace{1em} Isaac Zhang \hspace{1em} Thomas Kim \hspace{1em} Caleb Banister}
\vspace{0.4em} \\
LayerZero Labs}
\maketitle

\let\thefootnote\relax\footnotetext{Copyright \copyright{~2023 LayerZero Labs Ltd. All rights reserved.}}
\let\thefootnote\relax\footnotetext{(2024-01-23) Version 1.1: Fix typos.}
\let\thefootnote\relax\footnotetext{(2024-03-07) Version 1.2: Fix figure typo.}
\let\thefootnote\relax\footnotetext{(2025-05-13) Version 1.3: Fix typo in Table \ref{tab:endpoint-api}}

\begin{abstract}
In this paper, we present the first \emph{intrinsically secure} and \emph{semantically universal} omnichain interoperability protocol: LayerZero.
Utilizing an \emph{immutable} endpoint, append-only verification modules, and fully-configurable verification infrastructure, LayerZero provides the security, configurability, and extensibility necessary to achieve omnichain interoperability.
LayerZero enforces strict application-exclusive ownership of protocol security and cost through its novel trust-minimized \emph{modular security} framework which is designed to universally support all blockchains and use cases.
Omnichain applications (OApps) built on the LayerZero protocol achieve frictionless blockchain-agnostic interoperation through LayerZero's universal network semantics.
\end{abstract}
\section{Introduction}
\label{sec:introduction}

Blockchain interoperability represents an ever-growing challenge as the diversity of chains continues to grow, and the importance of connecting the fragmented blockchain landscape is increasing as applications seek to reach users across a progressively wider set of chains.
We present LayerZero, the first omnichain messaging protocol (\emph{OMP}) to achieve a fully-connected mesh network that is scalable to all blockchains and use cases.

In contrast to the monolithic security model of other cross-chain messaging services, the LayerZero protocol uses a novel modular security model to \emph{immutably} implement security.
This approach can still be extended to support new features and verification algorithms.
\emph{Intrinsic security} against censorship, replay attacks, denial of service, and in-place code modifications is designed into \emph{immutable} Endpoints.
Less fundamental \emph{extrinsic} aspects of security (e.g., signature schemes) are isolated into independently-immutable modules.
As a protocol, LayerZero is not bound to any infrastructure or blockchain; all components other than the endpoint can be interchanged and configured by applications built on LayerZero.

We illustrate the omnichain fully-connected mesh network in Figure~\ref{fig:omni-vs-cross}.
Each chain is directly connected to every other chain, and while the \emph{extrinsic security} (Section~\ref{sec:principles-intrinsic-security}) may be different for different chain pairs (illustrated by the colored solid lines), the guarantees of eventual, lossless, exactly-once packet delivery should be uniform and never change.

LayerZero's network channel semantics, including execution features, configuration semantics, censorship resistance, and failure model, are universal.
These universal semantics allows application developers to easily architect secure, chain agnostic omnichain applications (\emph{OApps}).

\begin{figure}
    \centering
    \includegraphics[width=\columnwidth]{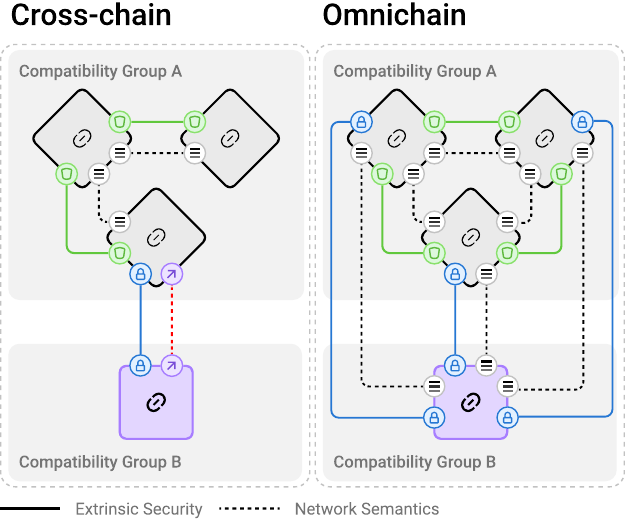}
    \caption{The omnichain fully-connected mesh network has universal network semantics for all connected chains and security specialized to each link.}
    \label{fig:omni-vs-cross}
    \vspace{-1em}
\end{figure}

The remainder of this paper is organized as follows.
First, Section~\ref{sec:principles} explains the overarching fundamental principles underlying the LayerZero protocol.
Section~\ref{sec:design-core} describes the protocol design and highlights how each component is architected for security.
Finally, Section~\ref{sec:extensions} presents examples of how \System{} can easily be extended to support a wide range of additional features in a blockchain-agnostic manner.
\section{Principles}
\label{sec:principles}

The responsibilities of an \MP{} can be condensed into two requirements: intrinsic security and universal semantics.
Existing messaging services fail to implement one or both of the above requirements and thus suffer from two fundamental deficiencies: monolithic security and overspecialization.
In the remainder of this section, we contextualize security and semantics within the cross-chain messaging paradigm, describe how existing cross-chain messaging systems fall short of these goals, and outline how LayerZero is designed from the ground-up to overcome these shortcomings.

\begin{table}[]
    \centering
    \begin{tabular}{r@{\,}l|l r}
        & Integrity Layer       & Failure model  \\
        \cline{2-3} \cline{2-3}
        \ldelim\{{7}{*}[\MP{}]  & Channel validity    & Packet censorship \\
        &                       & Packet replay \\
        &                       & Buggy updates \\
        &                       & Invalid reconfiguration \\
        &Channel liveness       & Denial of service \\
        &                       & Infrastructure health \\
        &                       & Administrator health \\
        \cline{2-3} \cline{2-3}
        
        &Data validity          & Cryptographic attack \\
        &                       & Malicious infrastructure \\
        &Data liveness          & Data loss on blockchain \\
    \end{tabular}
    \caption{We divide protocol integrity into four properties. Data liveness depends on the underlying blockchains and cannot be secured by the \MP{}.}
    \label{tab:integrity-layers}
\end{table}

\subsection{Security}
\label{sec:principles-intrinsic-security}
The first and most important requirement of \MP{}s is that they should be secure.
We divide security into \emph{intrinsic} and \emph{extrinsic} security, and while all messaging systems implement extrinsic security, few provide intrinsic security.
Intrinsic security refers to protocol-level invariants of lossless (censorship resistance), exactly-once (no replay), eventual (liveness) delivery.
Extrinsic security encompasses all other security properties, such as signature and verification algorithms.

Most existing messaging services have taken an ad-hoc approach to security, continuously updating a single monolithic end-to-end security model to accommodate chains as they are added to their network.
These services invariably utilize forced, in-place updates to a shared security model, and thus cannot provide long-term security invariants for OApps to build upon.
To provide long-term security invariants in LayerZero, we chose instead to modularize security and enforce strict immutability for all modules.

Table~\ref{tab:integrity-layers} illustrates how we divide protocol integrity into \emph{channel} and \emph{data} integrity.
Each of these integrity layers is further subdivided into validity and liveness properties.
We more formally define intrinsic security to cover channel validity and liveness, and \emph{extrinsic} security to cover data validity.
In this paper, we refer to the extrinsic security configuration as the \emph{Security Stack}.
Monolithic shared security systems force the same Security Stack on all applications, while isolated security systems allow a different Security Stack per OApp.

Intrinsic security can and should be universally secured based on first principles.
In contrast, optimal, trustless communication across blockchains is impossible, and the continuous advancement in verification algorithms and blockchain design necessitates \emph{extensibility} and \emph{configurability} of extrinsic security.
This is true even in special cases such as L1--L2 rollups; the possibility of hard forks necessitates L2 contract upgradability, thus making the L2 contract owner a trusted entity.

\begin{figure}
    \centering
    \includegraphics[width=1.00\columnwidth]{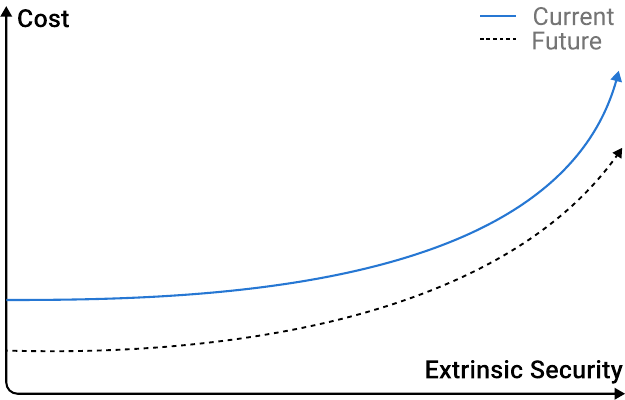}
    \caption{The pareto frontier of extrinsic security vs cost continually changes due to advancements in technology.}
    \label{fig:security-cost-tradeoff}
\end{figure}

It is infeasible to formally verify the extrinsic security of any nontrivial code given the many underlying layers of execution and the reliance of cryptographic algorithms on the computational intractability of NP problems.
As a result, the most practical measure of security is an economic one: a \emph{non-upgradable} smart contract's security is directly proportional to how many assets it has secured and for how long.
Thus, the implications of extrinsic security isolation are clear: \MP{}s must guarantee indefinite access to well-established, extrinsically secure code while still allowing protocol maintainers to extend the protocol.
The impossibility of trustless communication thus implies that extrinsic security exists on a constantly shifting pareto frontier (Figure~\ref{fig:security-cost-tradeoff}) and should be customizable to OApp-specific requirements.

To provide intrinsic security, the OMP must guarantee that an OApp's Security Stack only changes when the OApp owner \emph{opts in} to the change.
This implies that systems designed to allow \emph{in-place} code upgrades can \emph{never} be intrinsically secure.
Replaceability of existing code permits the permanent deprecation of well-established extrinsically secure code and the potential introduction of vulnerable code.
Current approaches to in-place secure upgrades involve careful testing and audits, but history has shown~\cite{nomad-hack,wormhole-vulnerability,thorchain-medium} that this process is not foolproof and can overlook severe vulnerabilities.
To guarantee long-term security invariants, \MP{}s must be architected to isolate each OApp's Security Stack from software updates and other OApps' configurations.

\subsection{Universal Semantics}
\label{sec:principles-universal-semantics}

The second requirement of \MP{}s is universal semantics, or the ability to extend and adapt the network primitive to all additional use cases and blockchains.

Execution semantics (i.e., feature logic) should be both chain-agnostic and sufficiently expressive to allow any OApp-required functionality.
A key insight we had when designing \System{} is that feature logic (execution) can be fully isolated from security (verification).
This not only simplified protocol development, but also eliminates concerns about impact to protocol security when designing and implementing \emph{execution features}.

The other aspect of universal semantics is universal compatibility of the \MP{} interface and network semantics with all existing and future blockchains.
The importance of semantic unification cannot be understated, as OApps \emph{cannot} scale if every additional blockchain in the network incurs significant engineering cost to accommodate different interfaces and network consistency models.
In practice, an \MP{} must have a unified interface, transmission semantics, and execution behavior regardless of the source and destination blockchain characteristics.

\begin{figure}[!ht]
    \centering
    \includegraphics[width=\columnwidth]{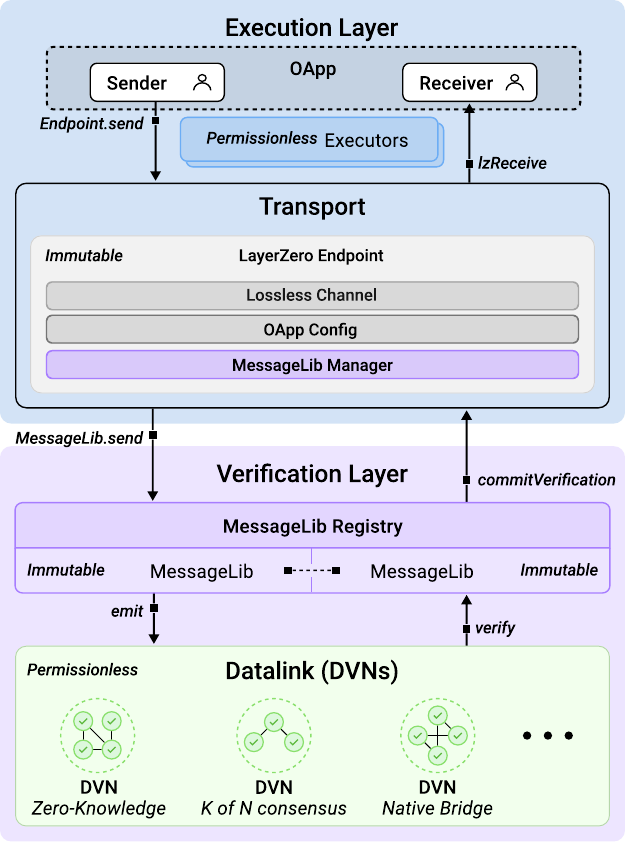}
    \caption{\System{} is divided into execution and verification layers. The verification layer securely transmits data between blockchains, and the execution layer interprets this data to form a secure, censorship resistant messaging channel.}
    \label{fig:layers}
\end{figure}

\section{Core protocol design}
\label{sec:design-core}

\begin{figure*}
    \centering
    \includegraphics[width=\textwidth]{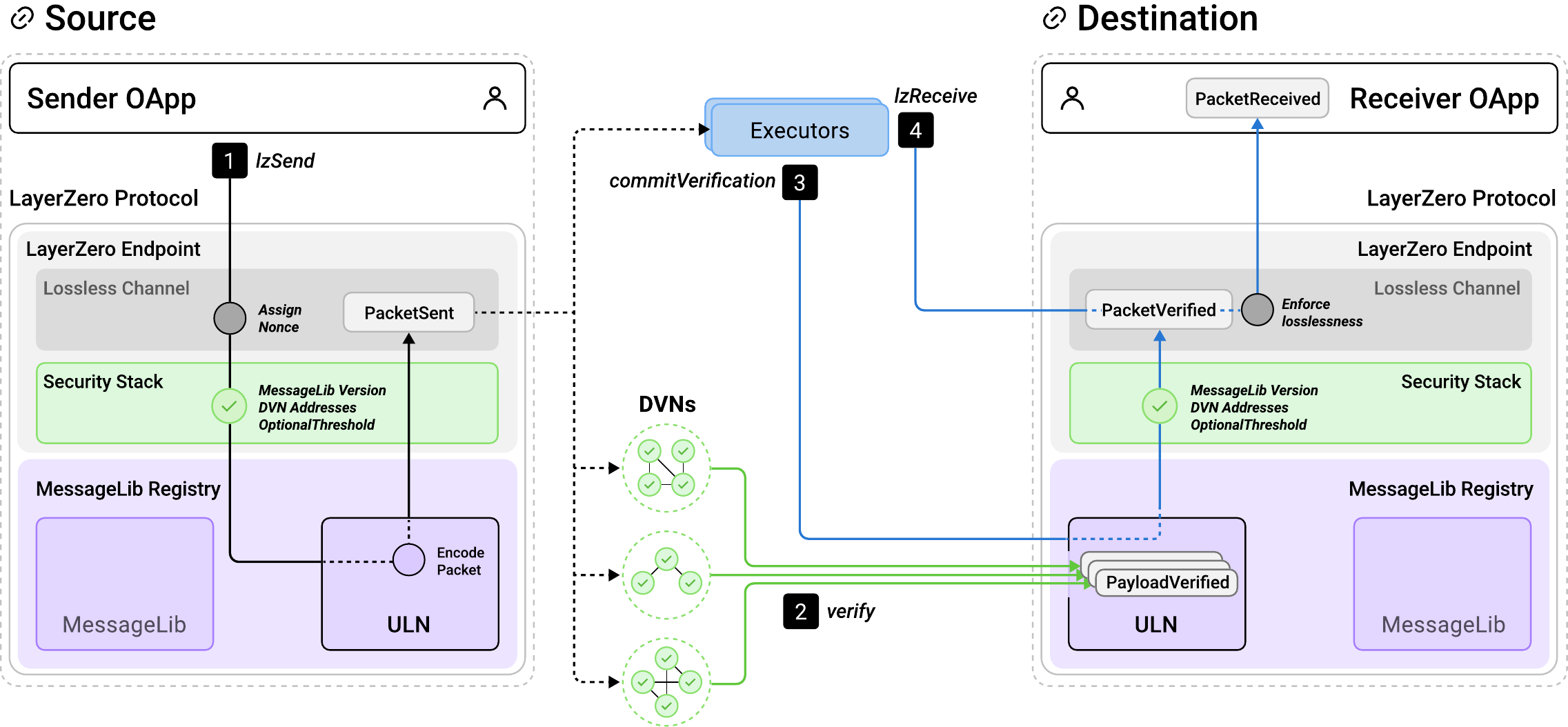}
    \caption{Steps to send a message using \System{}.}
    \label{fig:arch}
\end{figure*}

We divide \System{} into four components (Figure~\ref{fig:layers}): an immutable \emph{Endpoint} that implements censorship resistance, an append-only collection (\emph{MessageLib Registry}) of onchain verification modules (\emph{MessageLibs}), a permissionless set of Decentralized Verifier Networks (\emph{DVN}s) to verify data across blockchains, and permissionless \emph{\executor{}s} to execute feature logic in isolation to the cross-chain message verification context.
Tying the components together is the \emph{OApp Security Stack}, which defines the extrinsic security configuration of the protocol and is modifiable exclusively by the OApp owner.

\emph{Messages} in LayerZero are composed of a payload and routing information (path).
These messages are serialized into \emph{packets} before they are transmitted across the mesh network.
Packets are verified by the verification layer on the destination blockchain before they are \emph{committed} into the lossless channel.
The packets are then read from the channel and \emph{delivered} by \emph{executing} the \emph{lzReceive} callback on the destination OApp contract.

The LayerZero Endpoint secures channel validity through OApp-exclusive Security Stack ownership in conjunction with an immutable channel that implements censorship resistance, exactly-once delivery, and guaranteed liveness.
Endpoint immutability guarantees that no external entity or organization can ever forcibly change the security characteristics an OApp's Security Stack.

Individually immutable MessageLibs collectively form the MessageLib Registry, and each MessageLib is an extrinsically secure interface that verifies packet data integrity before allowing messages to be committed to the Endpoint.
Existing MessageLibs cannot be modified, thus making the MessageLib Registry append-only, and each Security Stack specifies exactly one MessageLib.
Security Stack ownership semantics in conjunction with MessageLib immutability enable applications to potentially use the same Security Stack \emph{forever}.

Each DVN is an aggregation of \emph{verifiers} that collectively verify the integrity of data shared between two independent blockchains.
DVNs can include both offchain and/or onchain components, and each Security Stack can theoretically include an unbounded number of DVNs.
The underlying DVN structure can leverage any verification mechanism, including but not limited to zero-knowledge, side chains, K-of-N consensus, and native bridges.
For brevity, we refer to the abstract collection of MessageLib, DVNs, and other hyperparameters as the Security Stack, and a serialized form of the Security Stack is assumed to be written to each chain.

Channel liveness (eventual delivery) is guaranteed through \emph{permissionless execution} in conjunction with \emph{Security Stack reconfiguration}.
Assuming the liveness of the source and destination blockchains, LayerZero channel liveness can only be (temporarily) compromised if (1) the DVNs in the Security Stack experience faults, or (2) the configured executor stops delivering messages.
If too many configured \verifier{}s stop verifying messages, the OApp can regain liveness by reconfiguring its Security Stack to use different \verifier{}s.
Packet delivery (execution) is permissionless, so any party willing to pay execution gas costs can deliver packets to restore channel liveness.

Interaction between \System{} components is minimized and standardized to reduce software bug surfaces (Figure~\ref{fig:layers}).
\System{}'s modularization and configurability also enables quick prototyping of the protocol on new chains.
Using a simple whitelist as a placeholder for MessageLib allows parallel development and testing of all components, expediting expansion of the mesh network to new chains.

\subsubsection{LayerZero packet transmission}
Before describing each component in detail, we present an overview of how packets are transmitted in LayerZero as shown in Figure~\ref{fig:arch}.
The \System{} mesh network is formed by the deployment by a protocol administrator of a \System{} Endpoint on each connected blockchain.
In this example, the OApp sends a LayerZero message from a \emph{sender} contract to a \emph{receiver} contract across the \System{} mesh network.
For illustration purposes, the MessageLib we use in this example is the \emph{Ultra Light Node}~\cite{layerzero-whitepaper} (Section~\ref{sec:design-ultra-light-node}).

During initial setup, the OApp configures its Security Stack on the \System{} Endpoint on the source and destination blockchains.
The MessageLib version configured in the Security Stack determines the \emph{packet version}.

In step \step{1}, the sender calls \texttt{lzSend} on the source chain \System{} Endpoint, specifying the message payload and the \emph{path}.
This path is associated with an independent censorship resistant channel, and is composed of the sender application address, the source Endpoint ID, the recipient application address, and the destination Endpoint ID.

The source Endpoint then assigns a gapless, monotonically-increasing nonce to the packet.
This nonce is concatenated with the path, then the result is hashed to calculate the global unique ID (GUID) of the packet.
This GUID is used by offchain and onchain \emph{workers} (e.g., executors, DVNs) to track the status of \System{} messages and trigger actions.

The source Endpoint reads the OApp Security Stack to determine the correct source MessageLib (ULN in this example) to encode the packet.
The source MessageLib processes the packet based on the configured Security Stack, rendering payment to the configured DVNs to verify the message on the destination MessageLib and optionally specified executors to trigger offchain actions.
These DVN and executor identifiers along with any relevant arguments are serialized by MessageLib into an unstructured byte array called \emph{Message Options}.
After the ULN encodes the packet and returns it to the Endpoint, the Endpoint emits the packet to conclude the LayerZero \texttt{send} transaction.

In step \step{2}, the configured \verifier{}s each independently \emph{verify} the packet on the destination MessageLib; for ULN, this constitutes storing the hash of the packet payload.
After a threshold of \verifier{}s verify the payload (see Section~\ref{sec:design-ultra-light-node}), a worker (e.g., executor, DVN, user) commits the packet to the Endpoint in step \step{3}.
The Endpoint checks that the payload verification reflects the OApp-configured Security Stack before committing to the lossless channel.

Finally, in step \step{4}, an \executor{} calls \texttt{lzReceive} on the committed message to execute the Receiver OApp logic on the packet.
Step \step{4} will revert to prevent censorship if the channel cannot guarantee lossless exactly-once delivery.

\begin{table}[]
    \begin{tabular}{r@{\,}r|cr}
     & Field name & Type \\
    \cline{2-3} \cline{2-3}
    \ldelim\{{7}{*}[Header] & Packet version & uint8 &\\
     & Nonce & uint64 &\\
     & Source Endpoint ID & uint32 &\rdelim\}{4}{*}[Path]\\
     & Sender & uint256 &\\
     & Destination Endpoint ID & uint32 &\\
     & Receiver & uint256 &\\
     & GUID & uint256 &\\
     \cline{2-3}
     & Message Payload & bytes[] &\\
    \end{tabular}
    \caption{\System{} packets are composed of a header and body. The header includes the packet version and path. The body is composed of the actual message payload. Packets are identified by their globally unique identifier (GUID).}
    \label{tab:packetlayout}
\end{table}

\subsection{\System{} Endpoint}
\label{sec:design-core-endpoint}

\begin{table*}[ht]
    \centering
    \begin{tabular}{c|l|l}
        \textbf{Routine}        & \textbf{Arguments}        & \textbf{Description} \\
        \hline
        \texttt{send}           &                           & Sends a message through \System{}. \\
                                & \texttt{path}             & The path of the channel through which to send the message. \\
                                & \texttt{payload}          & Data transmit to the receiver.\\
                                & \texttt{Message Options}  & Byte array of arguments to be interpreted by MessageLib (optional).\\
        \hline
        \texttt{getInboundNonce}&                           & Returns the largest nonce with all predecessors received.\\
                                & \texttt{path}             & The path corresponding to the message channel.\\
        \hline
        \texttt{skip}           &                           & Called by the receiver to skip verification and delivery of a nonce.\\
                                & \texttt{path}             & The path of the channel to skip a nonce on.\\
                                & \texttt{nonce}            & The nonce of the message to skip (must be the inbound nonce + 1).\\
        \hline
        \texttt{clear}          &                           & Called by the receiver to skip a nonce that has been verified.\\
                                & \texttt{path}             & The path to skip a nonce on.\\
                                & \texttt{guid}             & The GUID of the nonce to skip.\\
                                & \texttt{message}          & The contents of the message to skip.\\
        \hline
        \texttt{lzReceive}      &                           & Called by the \executor{} to receive a message from the channel.\\
                                & \texttt{path}             & The path of the channel to receive a message from.\\
                                & \texttt{nonce}            & Output parameter for the nonce of the received message. \\
                                & \texttt{GUID}             & The GUID of the received message. \\
                                & \texttt{message}          & The message to receive. \\
                                & \texttt{extraData}        & Any extra data requested by the receiver.\\
        \hline
        \texttt{nilify}         &                           & Called by the receiver to temporarily invalidate a nonce.\\
                                & \texttt{path}             & The path of the channel to nilify a message on.\\
                                & \texttt{nonce}            & The nonce of the packet to nilify.\\
                                & \texttt{payloadHash}      & The hash of the payload to nilify. \\
        \hline
        \texttt{burn}           &                           & Called by the receiver to delete and skip a nonce.\\
                                & \texttt{path}             & The path of the channel to burn a message on.\\
                                & \texttt{nonce}            & The nonce of the packet to burn.\\
                                & \texttt{payloadHash}      & The hash of the payload to burn. \\
    \end{tabular}
    \caption{\System{} core messaging API.}
    \label{tab:endpoint-api}
\end{table*}

The \System{} Endpoint, implemented as an immutable open-source smart contract and deployed in one or more instances per chain, provides a stable application-facing interface (Table~\ref{tab:endpoint-api}), the abstraction of a lossless network channel with exactly-once guaranteed delivery, and manages OApp Security Stacks.
The immutability of the \System{} Endpoint guarantees long-term channel validity by enforcing update isolation, configuration ownership, and channel integrity.
The Security Stack is key to \System{}'s channel liveness guarantee, as it mediates the trust--cost relationship between OApps and the permissionless set of \verifier{}s.

OApps call \texttt{send} on the Endpoint to queue a message to be sent through \System{}, specifying the \texttt{path} (Table~\ref{tab:packetlayout}), the message payload, and an optional byte array (Message Options) containing serialized options to be interpreted by MessageLib.
Message Options is purposely unstructured, improving extensibility as we discuss in Section~\ref{sec:extensions}.
The complement to \texttt{send} is \texttt{lzReceive}, which is executed on the destination chain to consume the message with the specified \texttt{GUID}.
On the destination chain, the Endpoint handles calls to \texttt{lzReceive} and \texttt{getInboundNonce}, enforcing lossless exactly-once delivery to protect the integrity of the message channel.
\texttt{lzReceive} delivers the verified payload of this message to the OApp, provided the message can be losslessly delivered.
\texttt{getInboundNonce} returns the highest losslessly deliverable nonce, computing the highest nonce such that all messages with preceding nonces have been verified, skipped, or delivered.
To handle erroneously sent messages or malicious packets, OApps either call \texttt{clear} to skip delivery of the packet in question or \texttt{skip} to skip both verification and delivery.

In addition to \texttt{clear} and \texttt{skip}, the Endpoint provides two convenience functions \texttt{nilify} and \texttt{burn}.
Nilify invalidates a verified packet, preventing the execution of this packet until a new packet is committed from MessageLib; this function can be used to proactively invalidate maliciously generated packets from compromised DVNs.
Burn is a convenience function to allow OApps to \texttt{clear} a packet without knowing the packet contents; this is useful if a faulty Security Stack commits an invalid hash to the endpoint, or if an OApp needs to clear a nilified nonce.

\begin{figure}[!ht]
    \centering
    \includegraphics[width=\columnwidth]{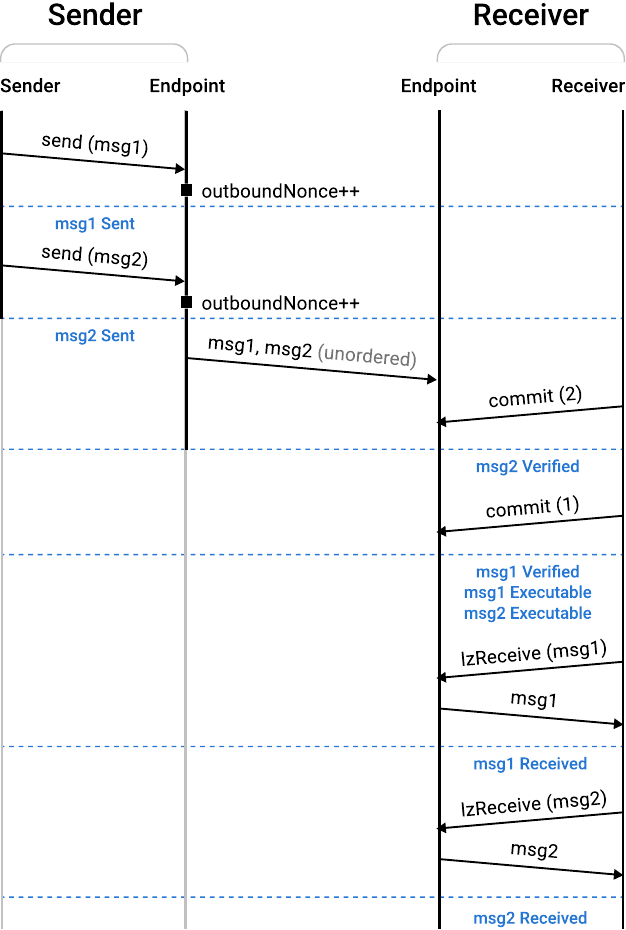}
    \caption{A packet is \emph{Sent} after source transaction increments the nonce, \emph{Verified} after it is committed into the Endpoint, and \emph{Received} after delivery (execution).}
    \label{fig:packet-lifecycle}
\end{figure}

\subsubsection{Out-of-order lossless delivery}

We lay out two non-negotiable consistency requirements for LayerZero's message channel: \emph{lossless} and \emph{exactly-once} delivery.
Censorship resistant channels \emph{must} be lossless, and exactly-once delivery is required to prevent replay attacks.
Both of these requirements are crucial for network integrity, and are guaranteed by the protocol provided the underlying blockchain is not faulty.

Channels in \System{} are necessarily separated and isolated by \texttt{path} (Table~\ref{tab:packetlayout}), as any lossless channel shared by two different OApps must sacrifice channel validity or liveness; an adversarial application can trivially deny liveness of a shared channel by refusing to verify a packet, but allowing other applications to forcibly skip the malicious OApp's packets constitutes censorship.
Each channel maintains a logical clock implemented by a gapless, strictly monotonically increasing positive integer nonce, and each message sent over the channel is assigned exactly one nonce.
On the destination Endpoint, each nonce is mapped to exactly one verified payload hash (Section~\ref{sec:design-msglib}), and the channel enforces that each delivered payload corresponds to the verified hash of the relevant nonce.
\System{} guarantees that the delivery of a packet implies all other packets on the same channel with lower nonces are delivered, deliverable, or skipped.

On a given channel, suppose two messages $m_k$ and $m_{k+1}$ are assigned positive integer nonces $k$ and $k+1$ respectively.
The gapless nonce guarantees that $\neg(\exists m_{j}~s.t.~m_{k} \rightarrow m_{j} \land m_{j} \rightarrow m_{k+1})$ where $\rightarrow$ is the happens-before relation.
More informally, there can be no packet with a nonce between $k$ and $k+1$.
This is the weakest possible, and by extension most flexible, condition for losslessness.
Stronger conditions (e.g., strictly in-order delivery) can be imposed on top of this abstraction if desired.

Censorship resistance is implemented by enforcing that no nonce can be delivered unless all previous nonces have been committed or skipped.
For example, a packet with nonce $N$ can only be delivered if all packets with nonce $1, \ldots, N-1$ are either committed or explicitly skipped by the receiver.
We term the largest nonce that can be executed to be the \emph{inbound nonce}.

Lossless and exactly-once delivery can be achieved using strictly in-order verification and execution, as demonstrated by Zarick et al.~\cite{layerzero-whitepaper}.
However, delivery order enforcement can result in artificial throughput limits on certain blockchains and complicates offchain infrastructure.
In \System{} we relax this ordering constraint, implementing out-of-order delivery that maintains channel integrity and does not introduce any additional onchain computational overhead.

The only efficient onchain implementation of an uncensorable channel with lossless, exactly-once, out-of-order delivery is to track the highest \emph{delivered} nonce, which we term the \emph{lazy inbound nonce}.
The lazy inbound nonce begins at zero, and packets can only be executed if all packets starting from the lazy inbound nonce until the packet nonce are verified.
Every time a packet is delivered (or skipped), the lazy inbound nonce is updated to the maximum of the current lazy inbound nonce and the nonce of the delivered packet.

The only other theoretical algorithm that can achieve efficient lossless and exactly-once delivery is updating the lazy inbound nonce upon \emph{verification} rather than execution.
This does not work in practice, because a single packet commit could result in an arbitrarily large update in the lazy inbound nonce.
For example, if nonce 2 through 1000 have been committed before nonce 1, the commit of nonce 1 must iterate through 1000 nonces to update the lazy inbound nonce.
This creates a situation where if the number of iterations exceeds the limits of the underlying blockchain, the corresponding channel will permanently lose liveness.

On the other hand, updating the lazy inbound nonce on execution can indeed run into computational limits, but permissionlessly retrying execution at a lower nonce will succeed.
It is possible for an uninterrupted sequence of undeliverable messages (i.e., messages that will always revert on execution) to cause temporary loss of channel liveness if the sequence is longer than the blockchain iteration limit.
This scenario can easily be rectified by the OApp owner calling \texttt{clear} (Table~\ref{tab:endpoint-api}) to skip execution of these undeliverable packets.

To enforce exactly-once delivery, we flag each packet after it is successfully received.
In LayerZero, this is implemented by deleting the verified hash of a packet from the lossless channel after it is delivered and disallowing verification of nonces less than or equal to the lazy inbound nonce.

We illustrate the lossless channel in Figure~\ref{fig:packet-lifecycle} through a simplified view of the lifecycle of a \System{} packet.
In this example, the OApp asynchronously sends two messages from the source chain to the destination chain, and each packet can be in one of three states: \emph{Sent}, \emph{Verified}, or \emph{Received}.
A packet is Sent after the source Endpoint assigns a nonce to the packet, after which the requested DVNs verify it on the destination MessageLib.
The packet transitions into the Verified state after an executor calls \texttt{commitVerification} (shown as ``commit'' in Figure~\ref{fig:packet-lifecycle}), which checks that the packet has been verified by the OApp Security Stack.

Once a packet and all preceding packets have been committed, the \executor{} calls \texttt{lzReceive} to deliver the packet and, barring reversion of the transaction, the packet transitions into the \emph{Received} state.
If one or more preceding nonces have not been committed, the lossless channel will revert to prevent censorship.

\begin{figure*}
    \centering
    \includegraphics[width=\textwidth]{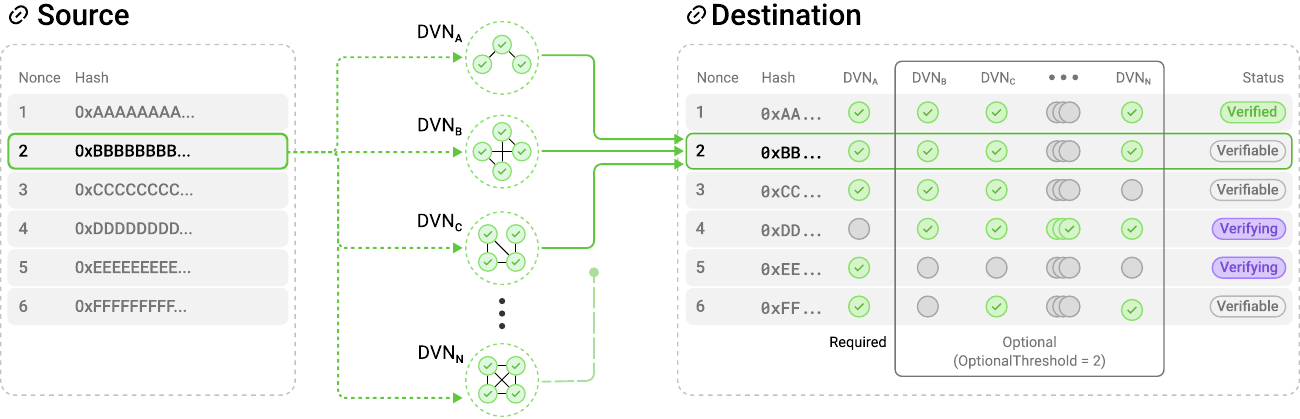}
    \caption{The Ultra Light Node enforces (onchain) the configured required DVNs, optional DVNs, and OptionalThreshold. Verification is neither lossless nor ordered, and messages can be committed to the channel as soon as the Security Stack is fulfilled.}
    \label{fig:uln-verification}
\end{figure*}

\subsection{MessageLib}
\label{sec:design-msglib}
The MessageLib Registry is a collection of MessageLibs, each of which are responsible for securely emitting packets on the source chain and verifying them on the destination MessageLib.
Each standalone MessageLib implements extrinsic security, necessitating adaptation to underlying environmental changes and precluding a fully immutable design of the MessageLib Registry.
MessageLib verifies the \emph{payload hash} of each packet, committing the verified payload hash to the endpoint after the extrinsic security requirement (e.g., DVN threshold) is fulfilled.

To provide extensibility for extrinsic security while protecting existing OApps against in-place updates, we structure the MessageLib Registry as an \emph{append-only} registry of immutable libraries, each of which can implement any arbitrary verification mechanism so long as it conforms to the protocol interface.
This design avoids the trap of validation lock-in that most messaging services fall into, and ensures that \System{} can be extended to take advantage of the most secure and efficient verification algorithms in every scenario.

This design may seem counterintuitive at first, as it precludes any in-place software updates and thus appears to prevent the protocol admin from addressing software bugs.
However, giving a single entity the power to unilaterally fix issues in-place also gives them the ability to introduce new vulnerabilities.
This in turn invalidates any long-term protocol security invariants, as a malicious change in the code can easily violate them.
Our decision to commit to immutable libraries in MessageLib presents higher barriers to introducing new code and bugfixes, but is crucial in realizing intrinsic security in \System{}.

We argue this append-only design is the \emph{only} way to implement intrinsic security without compromising extensibility.
In theory, a single non-upgradable MessageLib that is completely bug-free and perfectly optimized provides intrinsic end-to-end security, but is impractical even in this unrealistic scenario.
Changes in the execution environment (e.g., removal or addition of opcodes), consensus mechanism (e.g., validator election protocol), and evolving application preferences for verification algorithms necessitate protocol updates.
Any scheme that allows in-place modifications of MessageLibs is inherently not intrinsically secure; OApps are placing their trust in the developers, auditors, or governance structure of the protocol administrator, and there is no practical way to guarantee that the security of updated code will match or exceed the existing MessageLibs.
\MP{}s must allow extensions, but at the same time guarantee that the extrinsic security of previous versions is never impacted by these code additions.
Ergo, the only design that provides intrinsically secure updates is to append new versions of the codebase to an immutable registry of library versions.

Each MessageLib operates independently and handles the following tasks: (1) accept the message from the Endpoint, (2) encode and emit the packet (Table~\ref{tab:packetlayout}) to \verifier{}s and \executor{}s, paying any necessary fees, (3) verify the packet on the destination chain, and (4) commit the verified message to the destination Endpoint.
All other tasks are handled by \executor{}s, minimizing the code size of MessageLib and allowing easy addition of features through the implementation and operation of new \executor{}s.
No party, including the \System{} admin, is permitted to modify or remove libraries once they are added to the MessageLib registry.

Note that losslessness is enforced in the immutable endpoint (execution layer), \emph{not} in MessageLib (verification layer).
As we explain in Section~\ref{sec:design-ultra-light-node}, MessageLib can commit verified packet hashes into the endpoint out of order and with gaps.
However, packets cannot be consumed from the lossless channel if there are gaps in the sequence of verified packets.

\subsubsection{Ultra Light Node}
\label{sec:design-ultra-light-node}

The Ultra Light Node (ULN) is the baseline MessageLib included in every \System{} deployment, and allows the composition of up to 254 DVNs through customizable two-tier quorum semantics.
ULN implements the minimal set of fundamental features necessary for any verification algorithm and is thus universally compatible with all blockchains.
Each OApp Security Stack that is configured to use the ULN includes a set of \emph{required} \verifier{}s (X), \emph{optional} \verifier{}s, and a threshold ($OptionalThreshold$).
$X$ \verifier{}s are \emph{required}, and a packet can only be delivered if all $X$ required \verifier{}s and at least $OptionalThreshold$ optional \verifier{}s total have signed the corresponding payload hash.
After the necessary \verifier{} signatures have been aggregated on the ULN, the corresponding packet can be committed to the Endpoint.
The required \verifier{} model allows OApps to place a lower bound on the extrinsic security of the verification layer, as no message can be verified without a signature from the most secure \verifier{} in the required set.
This design delegates the majority of extrinsic security to the \verifier{}s while still enforcing the Security Stack onchain.

We illustrate an example of the Ultra Light Node verification semantics in Figure~\ref{fig:uln-verification}.
The OApp Security Stack includes required DVN (DVN$_A$) and $N-1$ optional DVNs (DVN$_B$, DVN$_C$, ...) with an OptionalThreshold of 1.
This gives DVN$_A$ ``veto'' power, and also requires at least one of the optional DVNs to verify the packet before it can be committed.
Nonce 1 has been committed to the messaging channel (\emph{Verified}).
Nonces 2, 3, and 6 are committable (verifiable) as the Security Stack has been fulfilled, but are not committed until an executor calls commitVerification.
Nonces 4 and 5 are not committable because they have not fulfilled the required verifier set and OptionalThreshold respectively.

This composable verification primitive gives OApps the ability to trade off cost and security, allows OApps to easily configure client diversity in their \verifier{} set, and minimizes the engineering cost of upgrading extrinsic security (no onchain code extensions).
The importance of client diversity cannot be understated, as even a non-compromised \verifier{} is subject to buggy code.

\subsubsection{MessageLib versioning and migration}
\label{sec:design-msglib-migration}
MessageLibs are identifiable through a unique ID paired with semantic (\texttt{major.minor}) version, and a message can only be sent between two Endpoints if both implement a MessageLib with the same major version.
Major versions determine packet serialization and deserialization compatibility, while minor versions are reserved for bugfixes and other non-breaking changes.
The packet version of each \System{} message is mapped to a MessageLib version, which \verifier{}s use to identify which MessageLib to submit packet verification to on the destination blockchain.
Each OApp Security Stack specifies the \texttt{sendLibrary} and \texttt{receiveLibrary} to use for each chain it spans.
The \texttt{sendLibrary} is the ID and version of the library to use when sending messages, and the \texttt{receiveLibrary} is the ID and version of the library to use when receiving messages.
This configurability enables OApps to customize Security Stack cost and security based on their individual needs.
For rapid prototyping, \System{} implements an opt-in mechanism to allow OApps to lazily resolve their Security Stack to the defaults chosen and maintained by the \System{} admin, but OApp owners are strongly encouraged to explicitly set their Security Stack for production applications.
Messages cannot be received on a path until an OApp has set their Security Stack or opted into the default Security Stack for that path.

The impossibility of coordinating atomic transactions over an asynchronous network~\cite{10.5555/139492.139503,cryptoeprint:2019/1128} necessitates a live migration protocol when reconfiguring the OApp Security Stack.

Upgrading to a MessageLib with the same major version, but different minor version (e.g., 1.1 $\rightarrow$ 1.2) is achieved by simply setting the \texttt{sendLibrary} and/or \texttt{receiveLibrary} to the desired version.
Migrating between MessageLibs with different major versions (e.g., 1.2 $\rightarrow$ 2.0) is more involved.
First, the OApp sets a grace period for the old \texttt{receiveLibrary} (1.2) during which \texttt{receiveLibrary} 1.2 can continue to receive messages even if the \texttt{sendLibrary} is configured to 2.0 by the OApp.
Next, the OApp sets the \texttt{sendLibrary} to the new version (2.0).
After this point all new messages carry the new packet version, but both \texttt{receiveLibrary} 1.2 and 2.0 can verify messages until the grace period elapses.
After the grace period elapses, only 2.0 is authorized to verify messages.
If the grace period ends before all version 1.2 in-flight messages are committed to the destination Endpoint, packet delivery will temporarily halt and the OApp must reconfigure \texttt{receiveLibrary} back to the previous version (1.2) to commit all in-flight messages before updating \texttt{receiveLibrary} to 2.0 again.
In this way, \System{} implements intrinsically secure, frictionless live migration between MessageLib versions.

\subsection{Decentralized Verifier Network}
\label{sec:design-core-trust}
The datalink in \System{} is designed on the fundamental observation that connecting two blockchains without the assumption of synchrony requires communication through one or more third parties~\cite{cryptoeprint:2019/1128}.
A consequence of the potentially-offchain nature of \verifier{}s is the impossibility of guaranteeing long-term immutability and availability, and as such a permissioned verification model is inherently unable to provide strong guarantees of channel liveness.
Thus, we have opted to implement a permissionless, configurable verification model in \System{}, where anyone can operate and permissionlessly integrate their own \verifier{} with \System{}.
Decentralized Verifier Networks, as the name suggests, are composed internally of a set of verifiers that collectively perform distributed consensus to safely and reliably read packet hashes from the source blockchain; this design importantly allows for client diversity within a single \verifier{}, minimizing the chance of a single faulty verifier causing protocol outages or errors.

\begin{figure}
    \centering
    \includegraphics[width=\columnwidth]{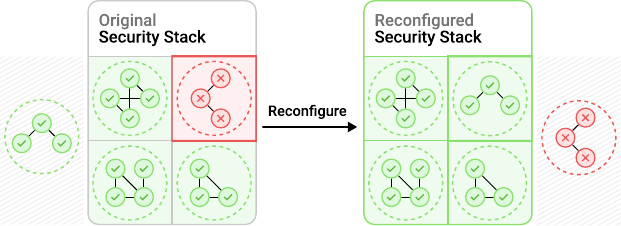}
    \caption{OApps can easily reconfigure their Security Stack to exclude faulty \verifier{}s.}
    \label{fig:risk-isolation}
\end{figure}

\begin{table*}
    \centering
    \begin{tabular}{l|l}
        Type & Structure \\
        \hline
         Execution gas & \lbrack~TYPE\_1, executionGas~\rbrack \\
         Gas and native drop & \lbrack~TYPE\_2, executionGas, nativedropAmount, receiverAddress~\rbrack  \\
         Composite & \lbrack~TYPE\_3, \lbrack~workerID, opType, length, command~\rbrack, ...~\rbrack \\
    \end{tabular}
    \caption{Message Options starts with a magic number to identify the options type, followed by type-specific options.
    Type 1 and 2 are specialized for setting execution gas limits and sending additional native gas tokens as part of an omnichain transaction respectively.
    Type 3 embeds arguments for an arbitrary set of offchain workers.}
    \label{tab:execution-options}
\end{table*}

This model overcomes two glaring shortcomings of other messaging services: shared security and finite fault tolerance.
Existing cross-chain messaging services provide a single shared security configuration that is shared across all clients, and this is suboptimal in two ways: it gives OApps no recourse if the protocol-specified verifier set is compromised or faulty, and it requires the majority of OApps to choose between an unnecessarily expensive or excessively risky security configuration for the task at hand.
Finite fault tolerance is a problem that plagues all messaging services with a permissioned verification model, as the failure of the entire finite-sized permissioned set of operating organization(s) results in the permanent, unrecoverable failure of the whole protocol.

Through permissionless operation of \verifier{}s, \System{} is able to provide a practically unbounded degree of fault tolerance.
Even if \emph{all} existing \verifier{}s lose liveness from software bugs, security breaches, natural disasters, and/or operational/governance concerns, OApp developers can operate their own \verifier{}s to continue operation of the protocol.
OApps can seamlessly reroute traffic through Security Stack reconfiguration, enabling recovery from compromised offchain infrastructure (Figure~\ref{fig:risk-isolation}).

\subsection{Executors}
\label{sec:design-executor}

Implementing and updating extrinsically secure code is resource-intensive due to stringent security testing and auditing requirements.
This engineering challenge stands in conflict with our goal of making \System{} easily extensible to support the needs of a wide variety of omnichain applications.
\System{} solves this problem by separating verification from execution; any code that is not security-critical is factored out into \emph{\executor{}s}, which are permissionless and isolated from the packet verification scope.
This separation between security-critical and ``feature'' code between MessageLib and \executor{}s respectively provides two main benefits.
First, it allows developers to use, implement, and compose feature extensions without considering security; the Endpoint prevents \executor{}s from delivering unverified or non-lossless messages, completely isolating packet execution from verification.
Second, it decouples security and liveness in \System{}, ensuring that a faulty \executor{} cannot unilaterally prevent message delivery.
By isolating the verification and execution layers in this manner, applications can easily debug which layer an error originated from.

When an OApp sends a LayerZero message, it specifies all offchain workers (e.g., executors, DVNs, etc.) and corresponding arguments through a MessageLib-interpreted byte array called Message Options.
The executors then wait for the Security Stack to verify the packet before taking action based on the commands encoded in Message Options.

We purposely left Message Options unstructured (see Section~\ref{sec:extensions}), as a more restrictive API only serves to limit the potential capabilities of future onchain and offchain workers.
At the protocol level, our primary objective is to provide the highest degree of extensibility, rather than creating an interface that is specialized to the needs of current applications.

\begin{figure}
    \centering
    \includegraphics[width=\columnwidth]{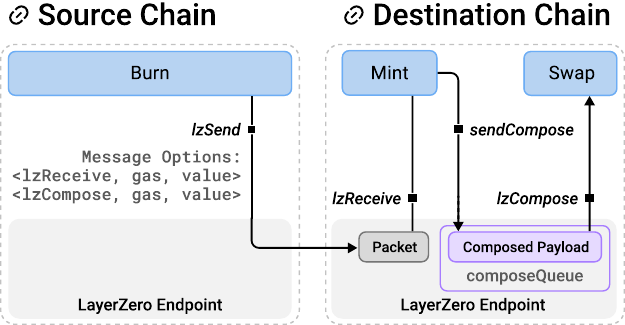}
    \caption{\texttt{lzCompose} enables chain-agnostic composition with liveness and safety closures.}
    \label{fig:lzcompose}
\end{figure}

The isolation of \executor{}s from any verification-related code indirectly improves channel validity, and permissionless execution directly improves channel liveness.
This design reduces the code footprint of MessageLib and, by extension, minimizes the potential to introduce attack surfaces into security-critical code.
In addition, permissionless operation of \executor{}s ensures that channel liveness can be recovered in the event of executor failure, and fully decouples the liveness of the protocol from any single organization or entity.
Once a message is verified by the Security Stack, anyone willing to pay the gas cost can permissionlessly execute the message.
This theoretically allows even end-users to manually trigger OApp recovery following \executor{} failure.
\section{Extensions}
\label{sec:extensions}
In this section, we illustrate the flexibility of \System{} through several examples of how the protocol can be extended with additional execution features.

\subsection{Message Options}
While there is no single standard format for serializing arguments into Message Options, we do not expect developers to write specialized code to support Message Options for every MessageLib.
To address this, \System{} currently defines three standardized formats for Message Options (see Table~\ref{tab:execution-options}) to facilitate backwards-compatibility between library versions.

Types 1 and 2 specify arguments for a single \executor{} to execute commonly-required functionality, while type 3 encodes a list of (workerID, option) tuples to allow for an arbitrary number of arguments passed to an arbitrary number of workers.
Any message delivered by a \executor{} has already been verified by the verification layer, allowing any number of \executor{}s to perform arbitrary actions without compromising message integrity.

\subsection{Semantically uniform composition}
\System{} defines a universally standardized interface for cross-chain composition: \texttt{lzCompose}.
Composing the destination chain delivery transaction with other contracts may seem trivial to those familiar only with EVM, which provides a native mechanism for arbitrary runtime dispatch.
However, even existing MoveVM-based blockchains do not natively support this feature, invalidating the universality of EVM-style runtime dispatch composition semantics.
When composing contracts, the receiver first stores a composed payload into the endpoint (\texttt{sendCompose}), after which it is retrieved from the ledger and passed to the composed callback by calling \texttt{lzCompose}.
This design, while superficially inefficient on EVM-based chains, unifies composition semantics across all blockchains.

\begin{figure}
    \centering
    \includegraphics[width=\columnwidth]{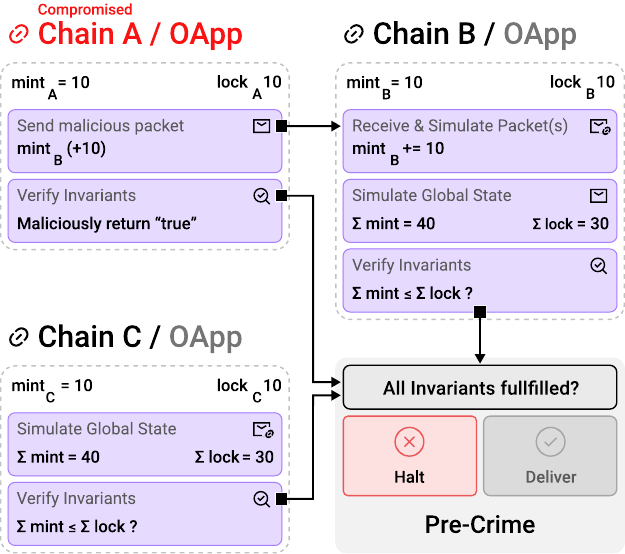}
    \caption{Pre-Crime rejects malicious and malformed messages by checking OApp-specified invariants.}
    \label{fig:precrime}
\end{figure}

\texttt{lzCompose} provides a semantically universal standard composition primitive that inherits the same only-once lossless execution semantics of \System{} messaging, and allows OApps to define a single application architecture that universally scales to all existing and future blockchains.
Figure~\ref{fig:lzcompose} illustrates an example of using \texttt{lzCompose} to bridge and swap a token in a single \System{} transaction.

The \texttt{lzCompose} primitive is a powerful tool for defining closures for data validity and channel liveness, isolating each composed contract from potential integrity violations by other contracts.
Once data is stored using \texttt{lzCompose}, the liveness and integrity are ``closed'', meaning errors in composed contracts can cause loss of liveness or data validity only \emph{within} the closure.
Isolation of composed contract faults to the closure scope greatly simplifies reasoning about potential attack surfaces such as reentrancy.

An additional benefit of \texttt{lzCompose} is a uniform interface for tracing and analysis of a potentially deep call stack for complex multihop omnichain transactions, giving OApp developers a powerful tool to tackle the potentially daunting task of debugging omnichain code.

\subsection{Application-level security}
\label{sec:precrime}

It is impossible to use the Message Options interface to extend the verification scope to include \emph{additional} data, but OApps can use it to detect and filter out verified-yet-malicious messages (e.g., buggy messages that would trigger OApp-level faults).
We introduce our novel offchain application-level security mechanism called \emph{Pre-Crime}, which provides an additional layer of application-specific packet filtering on top of the existing \System{} protocol.
Pre-Crime enables any subset of peers (i.e., some or all of an OApp's contracts) to enforce application security invariants after simulating the result of packet delivery.
The invariant check results are collated by an offchain worker, which halts delivery of the corresponding packet if any peer reports a violated invariant.

Figure~\ref{fig:precrime} illustrates the example of checking the total outstanding token count in a 3-chain token bridge.
To use Pre-Crime, the OApp encodes the \verifier{} address and Pre-Crime specific arguments in Message Options.
The OApp specifies the invariant that the total minted tokens across all chains $\sum{mint}$ is less than or equal to the total locked liquidity $\sum{lock}$.
Initially, $mint_A = mint_B = mint_C = 10$, and $lock_A = lock_B = lock_C = 10$.
Chain A is compromised and tries to request an additional mint of 10 tokens on chain B without locking any additional assets.
Pre-Crime detects this after checking token counts for all chains, isolating the security breach to a single chain (chain A).
The receiver can then skip the nonce if necessary by calling \texttt{skip} (Table~\ref{tab:endpoint-api}).
It is important to note that Pre-Crime does not add any additional \emph{protocol} security, and cannot protect data integrity (malicious DVNs or blockchain-level faults).

Pre-Crime and \texttt{lzCompose} are just two examples of execution features that are supported by LayerZero.
The flexibility of Message Options and the separation between verification and execution enables a LayerZero to be extended to a wide variety of execution features.
\section{Conclusion}
\label{sec:conclusion}
In this paper, we presented the design and implementation of the \System{} protocol.
\System{} provides intrinsically secure cross-chain messaging with universal semantics to enable a fully connected omnichain mesh network that connects all blockchains within and across compatibility groups.

By isolating intrinsic security from extrinsic security, \System{} guarantees long-term stability of channel integrity and gives OApps universal network semantics across the entire mesh network.
LayerZero's universal network semantics and intrinsic security guarantees enable secure chain-agnostic interoperation.

Our novel onchain verification module, MessageLib, implements extensible extrinsic security in an intrinsically secure manner.
Each OApp has exclusive permission to modify its Security Stack, which defines the extrinsic security (MessageLib, \verifier{}s) of their messaging channel.
The immutability of existing MessageLibs ensures that no entity, including the protocol administrator, can unilaterally compromise OApp security.

\System{}'s isolation of execution features from packet verification allows a near-unlimited degree of freedom to implement additional features without affecting security.
In addition, the separation of execution and verification in \System{} reduces engineering costs, attack surfaces, and improves overall protocol liveness.
Together, these components create a highly extensible protocol that can provide universal messaging semantics across existing and future blockchains.
\System{} is the protocol that brings consistency and simplicity to a landscape of scattered, ad-hoc messaging services, and lays the foundation for the fully-connected omnichain mesh network of the future.

{\footnotesize
\bibliographystyle{acm}
\bibliography{bib}}

\end{document}